\documentclass{aa}
\usepackage{natbib}
\usepackage{hyperref}
\hypersetup{colorlinks=true, citecolor=blue}
\usepackage{graphicx}
\usepackage{txfonts}
\bibpunct{(}{)}{;}{a}{}{,}

\begin{document}
  \title{Studying the spatially resolved Schmidt--Kennicutt law in interacting galaxies: the case of Arp~158}
  \author{M. Boquien\inst{1,2} \and U. Lisenfeld\inst{3} \and P.-A. Duc\inst{4} \and J. Braine\inst{5} \and F. Bournaud\inst{4} \and E. Brinks\inst{6} \and V. Charmandaris\inst{7,8,9} }
  \institute{University of Massachusetts, Department of Astronomy, LGRT-B 619E, Amherst, MA 01003, USA
  \and Laboratoire d'Astrophysique de Marseille, UMR 6110 CNRS, 38 rue F. Joliot-Curie, 13388, Marseille, France \email{mederic.boquien@oamp.fr}
  \and Departamento de F\'isica Te\'orica y del Cosmos, Universidad de Granada, Spain
  \and Laboratoire AIM, CEA/DSM - CNRS - Université Paris Diderot, DAPNIA/Service d'Astrophysique, CEA/Saclay, F-91191 Gif-sur-Yvette Cedex, France
  \and Laboratoire d'Astrophysique de Bordeaux, Universit\'{e} Bordeaux 1, Observatoire de Bordeaux, OASU, UMR 5804, CNRS/INSU, B.P. 89, Floirac F-33270, France
  \and Centre for Astrophysics Research, University of Hertfordshire, Hatfield AL10 9AB, UK
  \and University of Crete, Department of Physics, GR-71003, Heraklion, Greece
  \and IESL/Foundation for Research and Technology - Hellas, GR-71110, Heraklion, Greece
  \and Chercheur Associé\'e, Observatoire de Paris, F-75014, Paris, France
  }
  \date{30/06/2011}
  \abstract
  {Recent studies have shown that star formation in mergers does not seem to follow the same Schmidt--Kennicutt relation as in spiral disks, presenting a higher star formation rate (SFR) for a given gas column density.}
  {In this paper we study why and how different models of star formation arise. To do so we examine the process of star formation in the interacting system Arp~158 and its tidal debris.}
  {We perform an analysis of the properties of specific regions of interest in Arp~158 using observations tracing the atomic and the molecular gas, star formation, the stellar populations as well as optical spectroscopy to determine their exact nature and their metallicity. We also fit their spectral energy distribution with an evolutionary synthesis code. Finally, we compare star formation in these objects to star formation in the disks of spiral galaxies and mergers.}
  {Abundant molecular gas is found throughout the system and the tidal tails appear to have many young stars compared to their old stellar content. One of the nuclei is dominated by a starburst whereas the other is an active nucleus. We estimate the SFR throughout the systems using various tracers and find that most regions follow closely the Schmidt--Kennicutt relation seen in spiral galaxies with the exception of the nuclear starburst and the tip of one of the tails. We examine whether this diversity is due to uncertainties in the manner the SFR is determined or whether the conditions in the nuclear starburst region are such that it does not follow the same Schmidt--Kennicutt law as other regions.}
  {Observations of the interacting system Arp~158 provide the first evidence in a resolved fashion that different star--forming regions in a merger may be following different Schmidt--Kennicutt laws. This suggests that the physics of the interstellar medium at a scale no larger than 1~kpc, the size of the largest gravitational instabilities and the injection scale of turbulence, determines the origin of these laws.}
  \maketitle

\section{Introduction}

How atomic gas turns molecular and forms stars is one of the questions that is paramount to understanding the process of star formation in galaxies. This process plays a fundamental role in the transformation of baryonic matter and is the main driver of galaxy formation and evolution. It is well known that star--formation depends on the local physical conditions of the gas such as the pressure. It depends on the molecular gas column density by way of the Schmidt--Kennicutt law \citep{schmidt1959a,kennicutt1998b,kennicutt2007a,leroy2008a,bigiel2008a}. On the large scale it is, however, still not well understood how star formation depends on feedback, spiral density waves or shear, and more generally on the global environment such as in the case of mergers \citep{barnes2004a,chien2010a,teyssier2010a,bournaud2011a}.

One way to test for the effect of the environment is to use star--forming regions located in collisional debris which are the offsprings of galaxy interactions. Indeed, depending on the parameters of the collision (relative velocity, impact factor, prograde versus retrograde, etc.), a varying amount of gas and stars can be stripped from the parent galaxies and injected into the intergalactic medium. In addition to atomic gas pulled from the parent galaxies, these collisional debris actually contain surprisingly large amounts of molecular gas formed in--situ \citep{braine2000a,braine2001a,lisenfeld2002a,lisenfeld2004a,petitpas2005a,duc2007a}. As the gas subsequently collapses, star--forming regions are created with masses ranging from a few hundred solar masses, creating OB associations, the ``emission line dots'' \citep{gerhard2002a,yoshida2002a,sakai2002a,weilbacher2003a,ryan2004a,mendes2004a,cortese2006a,werk2008a,werk2010a} to objects as massive as dwarf galaxies named Tidal Dwarf Galaxies \citep[TDGs,][]{duc1995a,duc1998a,duc2000a,duc2007a,hancock2007a,hancock2009a}. These objects, even though formed from material once pertaining to their parent galaxies, have a radically different and simpler environment. Recently, \cite{boquien2007a,boquien2009a} showed that star--formation tracers such as ultraviolet, mid--infrared, and H$\alpha$, are as reliable in intergalactic star--forming regions as they are in spiral galaxies. \cite{braine2001a} showed that the depletion timescale of the molecular gas is similar to that of spiral galaxies even though the collision debris they studied have the luminosity, mass and colour of dwarf galaxies. As collision debris probably have a conversion factor close to that of spiral galaxies with a solar neighbourhood metallicity, this eliminates a major source of uncertainty.

There are still many open questions regarding star--formation in collision debris and how the process compares to star formation in spiral disks or in other environments. It has been observed that star formation is particularly efficient in ultraluminous infrared galaxies (ULIRGs) deduced from a high L(TIR)/L(CO) \citep{solomon2005a}, as well as in galaxies that have a subsolar metallicity \citep{leroy2006a,gardan2007a,gratier2010a,braine2010b}. Conversely, other systems like intergalactic collisional gas bridges contain molecular gas but form stars inefficiently, such as the UGC 813/6 pair \citep{braine2004b}. Recent results by \cite{daddi2010a} and \cite{genzel2010a} showed that starbursts and mergers follow different modes of star formation compared to more quiescent spiral galaxies. Starbursts appear to be forming stars at a higher rate than spiral galaxies for the same gas column density. However, the discrepancy between the 2 regimes  is increased by the use of a different X$_\mathrm{CO}$ conversion factor between star--forming galaxies and starbursting ones. Can we observe such a difference within a single interacting system, hence retrieving this result in a resolved fashion? Are there deviations in the Schmidt--Kennicutt law between intergalactic star formation regions and what is seen in galactic disks \citep{bigiel2008a}?

To address these questions we compare star formation in different parts of a single interacting system, Arp~158, and its tidal debris. This system is particularly suited for this study as it presents evidence of extended star formation. Using an homogeneous dataset on a single system means there are no internal calibration or methodological differences for the study of star formation in the interacting galaxies and their tidal debris. Arp~158 is an intermediate--stage merger in the \cite{toomre1977a} sequence, with the disks progressively merging but still with 2 clearly separated nuclei. With a recession velocity $cz=4758$~km~s$^{-1}$, Arp~158 is located at a luminosity distance of 62.1~Mpc\footnote{Value obtained from NED, the NASA Extragalactic Database operated by the Jet Propulsion Laboratory, California Institute of Technology, under contract with the National Aeronautics and Space Administration.}, assuming $\mathrm{H_0}=73$~km~s$^{-1}$~Mpc~$^{-1}$, $\Omega_\mathrm{m}=0.27$ and $\Omega_\mathrm{\Lambda}=0.73$, which corresponds to a scale of 292~pc/arcsec. To carry out this work we use multi--wavelength data, ranging from the far--ultraviolet to the radio, to trace star formation, the stellar populations and the atomic gas, also encompassing observations of the molecular component. Optically, the closely interacting galaxies have a boxy shape. They present a bar with 3 nearly aligned bright clumps. The nature of one of them, galactic nuclei or a foreground star, is debated in the literature \citep{chincarini1973a,dahari1985a}. Prominent dust lanes are visible North of the bar. Two tidal tails can be seen, containing blue concentrations at their tips. The longer one, oriented towards the East--South--East, is a prime Tidal Dwarf Galaxy candidate. The disks of the galaxies seem to be rather edge--on but their morphology is strongly disturbed making a precise assessment difficult. Radio 21~cm observations have shown that the entire system contains $6.5\times10^9$~M$_\odot$ of HI, with column density peaks at the tip of one of the tidal arms and towards the western nucleus \citep{iyer2004a}, providing a large reservoir of gas to fuel star formation throughout the system which yields an infrared luminosity of $\mathrm{3.1\times10^{10}~L_\odot}$ (Sec.~\ref{ssec:SFR}).

In Sec.~\ref{sec:observations} we present the multi--wavelength observations, followed by the results in Sec.~\ref{sec:results} which we discuss in Sec.~\ref{sec:discussion} before concluding in Sec.~\ref{sec:conclusion}.

\section {Observations and data reduction}
\label{sec:observations}

In this section we present the set of multi--wavelength data we are using to carry out this study. For an easier description of the observations we have labelled different regions of interest in the system. These regions have been selected from the inspection of multi--wavelength data and concern the most interesting morphological features. The main characteristics of each object are briefly described in Table \ref{tab:regions} and they are labelled on images of the system in Fig.~\ref{fig:system-image}.

\begin{table*}[!htbp]
\caption{Regions of interest\label{tab:regions}}
\begin{tabular}{lccl}
\hline
Label&$\alpha$&$\delta$&Comment\\\hline
NE&01:25:22.334&+34:01:32.39&Eastern nucleus\\
NC&01:25:20.734&+34:01:29.84&Central nucleus\\
NW&01:25:19.467&+34:01:30.08&Western object: star or nucleus\\
C1&01:25:20.447&+34:01:25.35&Optical faint blue region at the base of the southern arm\\
C2&01:25:16.830&+34:01:37.13&Several compact blue regions to the West, in the northern arm\\
C3&01:25:18.670&+34:01:36.30&Several compact blue regions in the northern arm\\
C4&01:25:21.893&+34:01:21.54&Faint compact blue regions in the southern arm\\
C5&01:25:25.644&+34:01:07.12&Clumpy compact blue regions in the TDG candidate to the East\\
\hline
\end{tabular}
\end{table*}

\subsection{CO}
\label{ssec:CO}

To trace the molecular gas content of the system we have carried out CO observations with the 30 metre millimetre--wave telescope on Pico Veleta (Spain) run by the Institut de Radio--Astronomie Millim\'etrique (IRAM). In November 2006 we observed positions C5 and NC with the A and B receivers. The observations of the positions NC, NE, C2 and C3  were done in January 2010 with the EMIR receiver. The CO(1--0) and CO(2--1) transitions at 115 GHz and 230 GHz, respectively, were observed simultaneously and in both polarisations. The redshifted frequencies were 113.4547 and 226.905 GHz, corresponding to a central velocity of $cz = 4800$~km~s$^{-1}$. For the November 2006 observations, we used the 512 $\times$ 1 MHz filterbanks at 3~mm, one for each polarisation, and the two 256 $\times$ 4 MHz filterbanks for the two polarisations at 1 mm, yielding instantaneous bandwidths of 1300~km~s$^{-1}$ in all transitions. The observations in January 2010 used the autocorrelator Wilma with a resolution of 2 MHz and for both the 90 GHz and 230 GHz band. The bandwidth was over 10000~km~s$^{-1}$ for the CO(1--0) line and over 5000~km~s$^{-1}$ for the CO(2--1) line. All the  observations were carried out in good weather conditions, the mean system temperatures being for the CO(1--0) transition 190 K (November 2006), and 245 K (January 2010), and for the CO(2--1) transition 330 K (November 2006), and 240 K (January 2010) on the T$_A ^*$ scale. The main beam efficiencies at Pico Veleta were taken to be at 115 GHz (0.74 and 0.77 for the A+B receivers and EMIR respectively), and at 230 GHz (0.54 and 0.58 for the A+B receivers and EMIR respectively). The halfpower beamwidths are about $\sim22$ and $\sim11$ arcseconds, corresponding to physical sizes 6.4~kpc and 3.2~kpc. All observations were done in wobbler--switching mode, with a throw in azimuth between 150 and 220 arcseconds. Pointing was checked on a nearby quasar roughly every 90 minutes. 

Data reduction was straightforward: the spectra for each position were averaged. In most cases only zero--order baselines (i.e. continuum levels) were subtracted to obtain the final spectra, just in a few spectra a linear baseline had to be subtracted. Position NC was observed during both observing runs and allowed us to check the relative calibration. We found that the velocity integrated intensities were within 10\% for CO(1--0) and CO(2--1). This constitutes the main source of uncertainty for the 3 brightest regions in CO. The characteristics of the lines are listed in Table~\ref{tab:CO}.

\begin{table*}[!htbp]
\caption{Characteristics of the CO lines for the 5 pointings.\label{tab:CO}}
\begin{tabular}{llllllllll}
\hline
Region&I$_\mathrm{CO(1-0)}$&v$_\mathrm{CO(1-0)}$&$\Delta$v$_\mathrm{CO(1-0)}$&I$_\mathrm{CO(2-1)}$&v$_\mathrm{CO(2-1)}$&$\Delta$v$_\mathrm{CO(2-1)}$\\
&(K~km~s$^{-1}$)&(km~s$^{-1}$)&(km~s$^{-1}$)&(K~km~s$^{-1}$)&(km~s$^{-1}$)&(km~s$^{-1}$)\\\hline
NC&$10.08\pm0.43$&4791&429&$9.0\pm0.55$&4825&277\\
NE&$5.87\pm0.38$&4900&248&$7.9\pm0.51$&4911&238\\
C2&$0.38\pm0.09$&4523&146&$<0.62$&4523&146\\
C3&$3.60\pm0.20$&4638&225&$3.07\pm0.13$&4589&128\\
C5&$0.44\pm0.05$&4955&100&$0.61\pm0.07$&4920&105\\\hline
\end{tabular}
\tablefoot{The velocity integrated intensities are in K~km~s$^{-1}$ in T$_\mathrm{mb}$ scale and velocities are in km~s$^{-1}$. The velocity width is taken from the 0--level of the spectra. The errors are 1--$\sigma$ errors derived from the rms noise of the spectra. A 10\% calibration error needs to be considered in addition.}
\end{table*}

\subsection{HI}
\label{ssec:HI}
The HI data cube was obtained from the literature \citep{iyer2004a}. Observations of the atomic hydrogen were performed at the VLA (Very Large Array) in D configuration between 1999-05-14 and 1999-05-17 for a total of 5~h. The velocity resolution is 12.3~km~s$^{-1}$ and the final map resolution is 43.3\arcsec$\times$42.2\arcsec\ corresponding to a physical size of 12.3~kpc$\times$12.6~kpc. Data reduction is described in detail in \cite{iyer2004a}.

\subsection{Ultraviolet, optical and infrared imaging}

Ultraviolet data have been obtained from the public archives of the GALEX space observatory. Observations were carried out on 2005--11--17 in the FUV (far ultraviolet; $\lambda_{\textrm{eff}}$=151~nm) and NUV (near ultraviolet; $\lambda_{\textrm{eff}}$=227~nm) bands in the context of the Guest Investigator programme GI2-019 (PI: Koratkar). Arp~158 is located $\sim3$\arcmin\ from the edge of the 1.24$^\circ$ field-of-view which is centered on NGC~507. The exposure time in FUV (NUV) is 3230~s (3398~s) leading to a sensitivity of 38~nJy~px$^{-1}$ (resp. 26~nJy~px$^{-1}$). The point spread function has a FWHM (full width half maximum) size of $\sim5$\arcsec\ corresponding to a physical size 1.5~kpc.

Shallow optical images have been obtained from SDSS (Sloan Digital Sky Survey) release 7 \citep{abazajian2009a}, tile 4829 in $u'$, $g'$, $r'$, $i'$ and $z'$ bands. Observations were carried out on 2004--09--15. Standard SDSS calibrations have been applied.

Shallow near--infrared images of the system were obtained from the 2MASS \citep[2 Micron All Sky Survey,][]{skrutskie2006a} archives. They are used exclusively for visual inspection.

Finally, Arp~158 was observed by our team in the 24~$\mu$m band on 2009--02--07 using the Spitzer Space Observatory, programme ID 50191. The final exposure time per pixel is 10~s. The frames were processed by the standard Spitzer pipeline version 18.5.0. The processed image presents a smooth gradient in the background which was eliminated subtracting on each row the median of the sky.

\begin{figure*}[!http]
\center
\includegraphics[width=\textwidth]{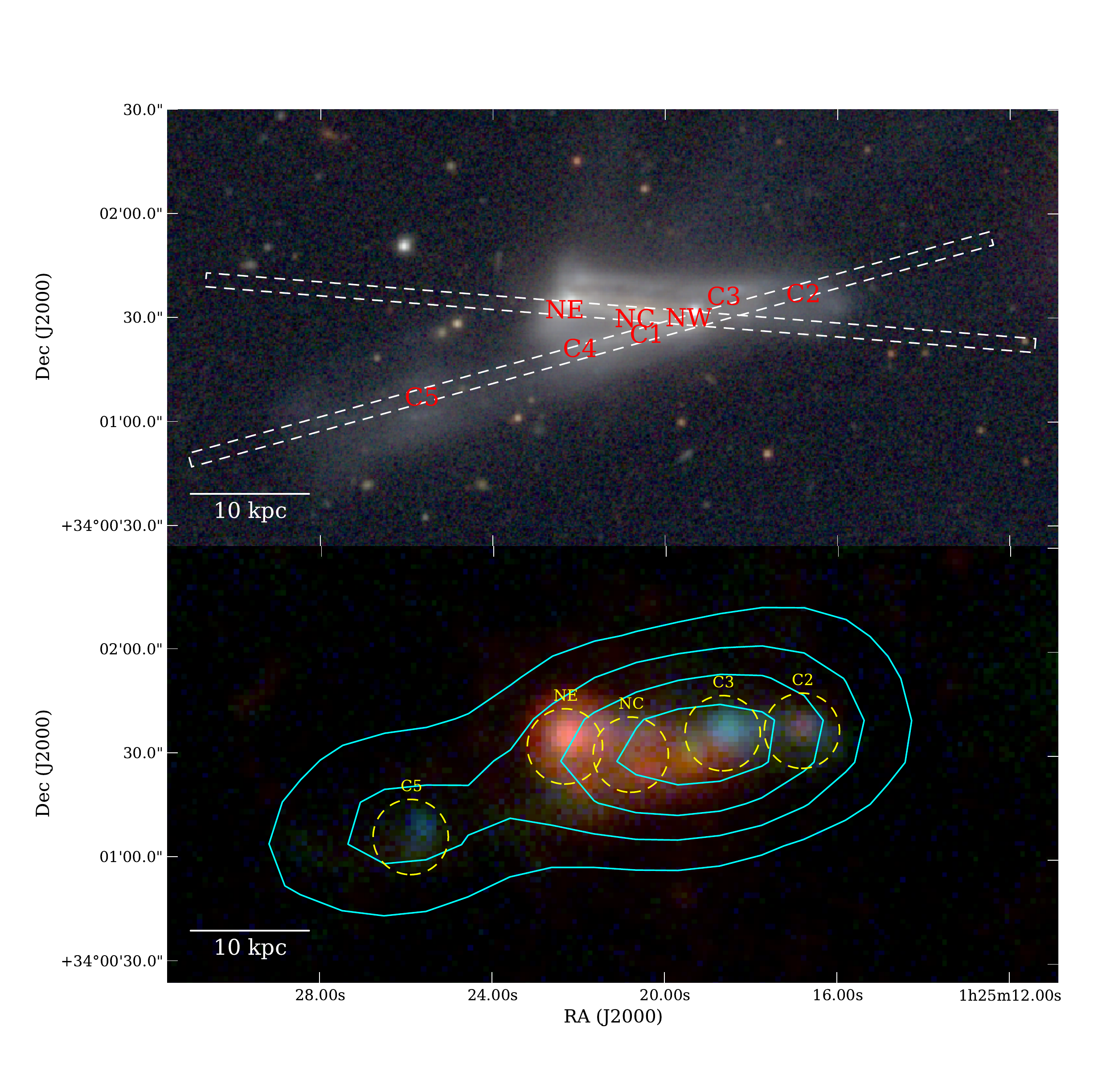}
\caption{Top: optical image in $g'$ (blue channel), $r'$ (green channel) and $i'$ (red channel) bands. The regions of interest listed in Table \ref{tab:regions} are indicated with a red label. The white dashed rectangles represent the two long slits used for optical spectroscopy. Bottom: FUV (blue channel), NUV (green channel) and 24~$\mu$m (red channel). The contours in cyan trace the HI column density. The yellow dashed circles represent the CO(1--0) beams and the area in which fluxes have been measured. Due to the large CO beams, not all regions of interest can have their CO emission detected separately.\label{fig:system-image}}
\end{figure*}

Images of the system combining various wavelengths are presented in Fig.~\ref{fig:system-image} along with the regions of interest identified.

Fluxes were measured in all images within circular apertures matching the CO(1--0) beam (about 22\arcsec\ for the half--power beamwidth, Sect. \ref{ssec:CO} and Fig.~\ref{fig:system-image}) using the {\sc phot} procedure from {\sc iraf}. The background was taken as the mean of the median pixel value in several positions around the interacting system. The same background was subtracted for all apertures. Recent studies of nearby galaxies have shown the importance of background subtraction to derive the star--formation laws \citep{blanc2009a,rahman2011a,liu2011a}. However, this is most important for regions in galaxies forming stars at a low rate that can be resolved only in nearby galaxies. The distance of Arp~158, 62.1~Mpc, and the size of the regions selected limits the bias introduced by the background subtraction. The uncertainty on the background level is taken as the quadratic mean of the standard deviation of the background level measurements and the mean of the standard deviation of pixels within apertures in which the background is measured. Foreground extinction from the Milky Way was corrected in the UV and in the optical using the \cite{cardelli1989a} extinction law assuming $\mathrm{E(B-V)}=0.054$ from NED. The foreground  extinction--corrected fluxes from FUV to 24~$\mu$m are presented in Table~\ref{tab:fluxes}.

\begin{table*}[!htbp]
\caption{UV, optical and IR fluxes in the 5 CO pointing beams.\label{tab:fluxes}}
\begin{tabular}{ccccccccc}
\hline
Region & FUV & NUV & $u$ & $g$ & $r$ & $i$ & $z$ & 24~$\mu$m\\
&($\mu$Jy)&($\mu$Jy)&(mJy)&(mJy)&(mJy)&(mJy)&(mJy)&(mJy)\\
\hline
NC&$76\pm4.1$&$189\pm9.9$&$1.13\pm0.03$&$4.34\pm0.09$&$8.76\pm0.18$&$13.1\pm0.27$&$17.7\pm0.39$&$57.3\pm2.3$\\
NE&$101\pm5.3$&$270\pm13.8$&$1.51\pm0.03$&$5.54\pm0.11$&$10.3\pm0.21$&$14.0\pm0.28$&$17.5\pm0.38$&$139.0\pm5.6$\\
C2&$52\pm3.0$&$94\pm5.6$&$0.29\pm0.01$&$0.82\pm0.02$&$1.24\pm0.03$&$1.56\pm0.05$&$1.72\pm0.16$&$6.8\pm0.5$\\
C3&$132\pm6.8$&$245\pm12.6$&$1.58\pm0.03$&$4.88\pm0.10$&$7.46\pm0.15$&$9.18\pm0.19$&$10.3\pm0.26$&$18.6\pm0.8$\\
C5&$31\pm2.1$&$55\pm4.1$&$0.17\pm0.01$&$0.46\pm0.02$&$0.71\pm0.02$&$0.93\pm0.05$&$1.0\pm0.16$&$2.1\pm0.4$\\
\hline
\end{tabular}
\tablefoot{The 1--$\sigma$ error bars take into account uncertainties on the absolute flux calibration and on the background level. Note that the errors are generally dominated by the uncertainty on the flux calibration.}
\end{table*}

\subsection{Optical spectroscopy}

\begin{figure*}[!htbp]
\center
\includegraphics[width=\columnwidth]{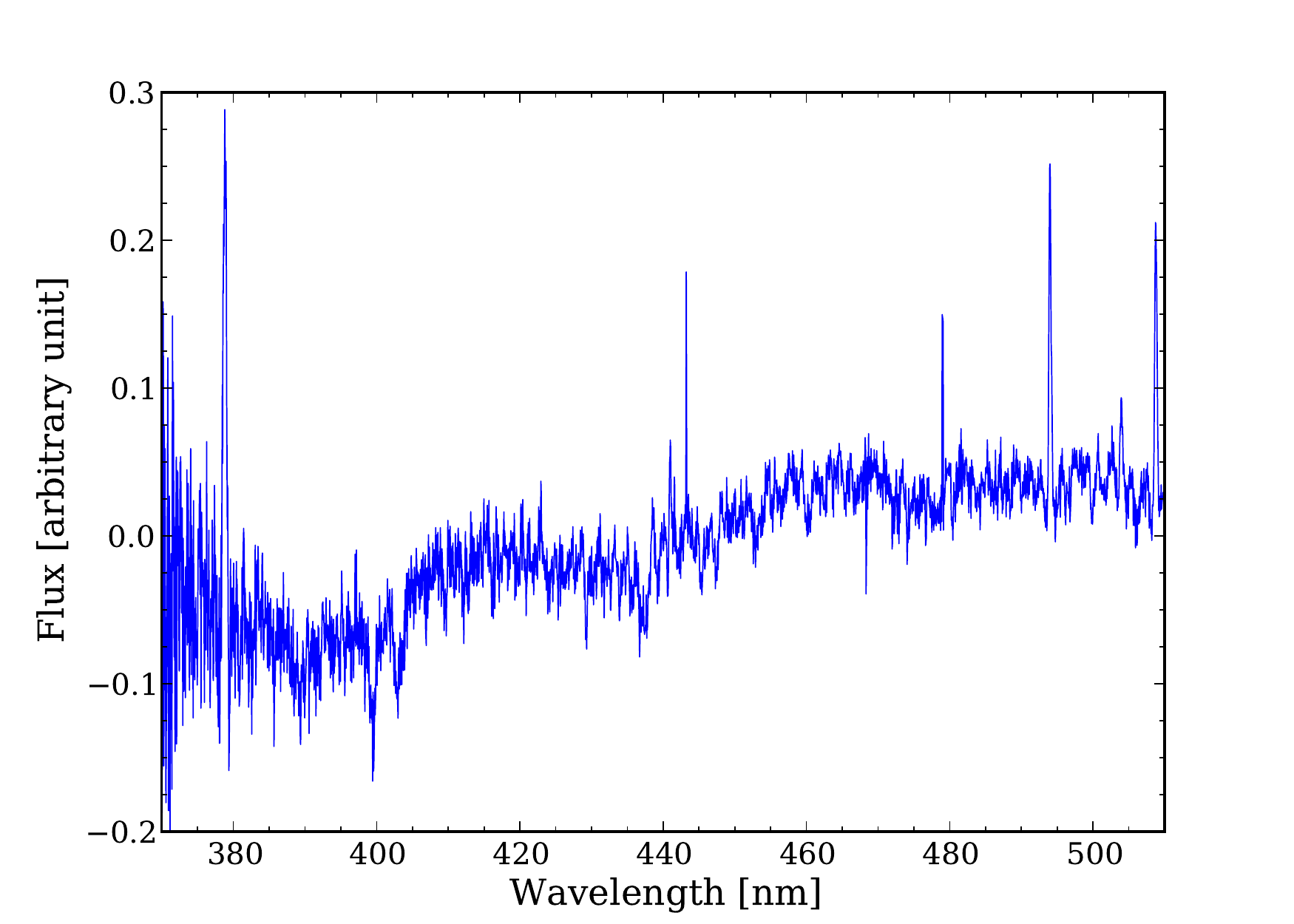}
\includegraphics[width=\columnwidth]{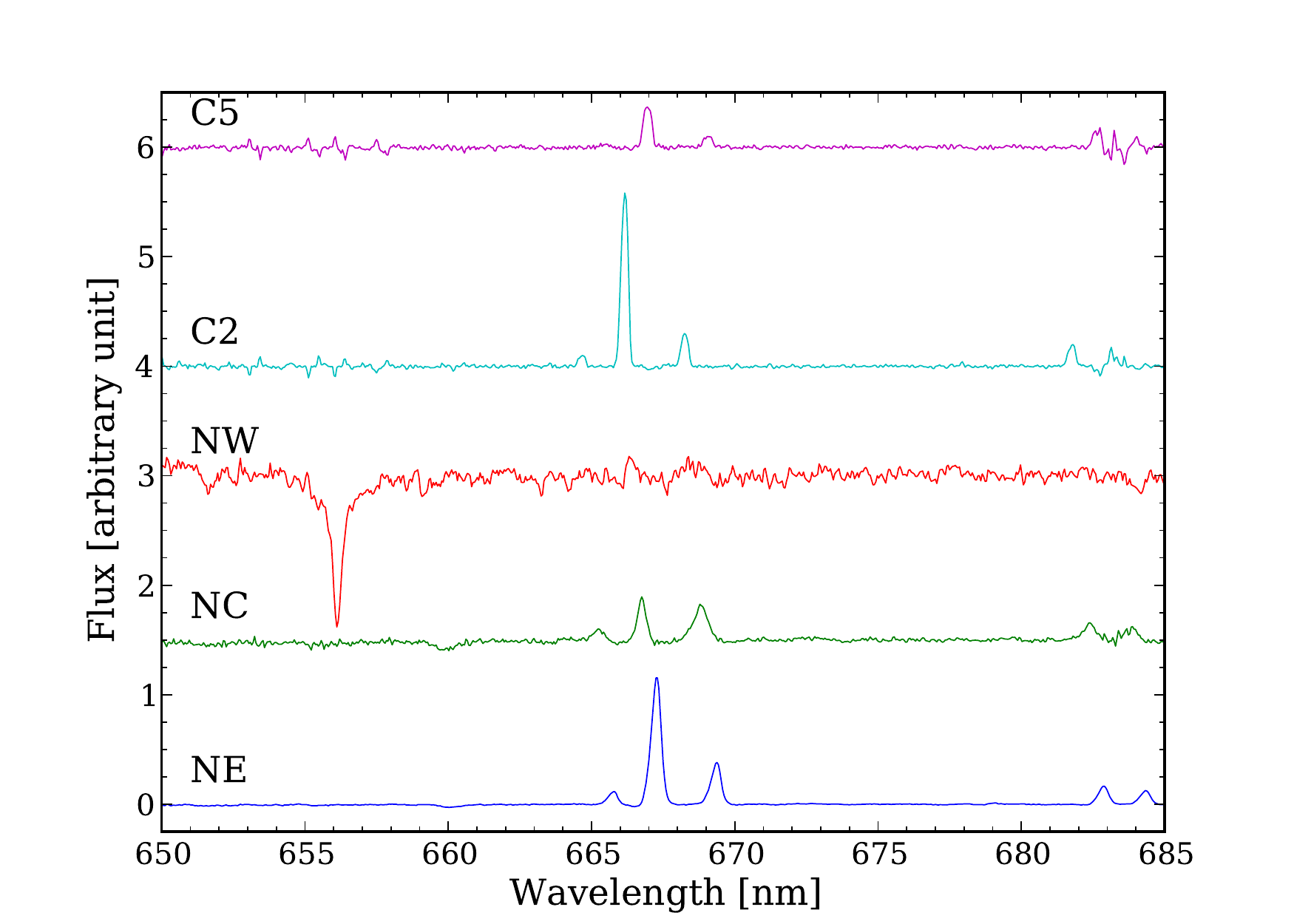}
\caption{Left: blue part of the spectrum of the NE clump. Several lines are clearly detected: $[\mathrm{OII}]_{372.7}$, H$\gamma$, $[\mathrm{OIII}]_{436.3}$, H$\beta$, $[\mathrm{OIII}]_{495.9}$, and $[\mathrm{OIII}]_{500.7}$. Unfortunately the blue spectrum for other clumps is not deep enough to detect these emission lines. Right: spectra of clumps NE, NC, NW, C2, and C5. The spectra have been offset to distinguish them more easily. The spectra of clumps NC, NW, C2, and C5 are multiplied by 5 as they are intrinsically much fainter than the spectrum of NE. Except for NW, several lines are clearly detected: $[\mathrm{NII}]_{654.8}$, H$\alpha$, $[\mathrm{NII}]_{658.4}$, $[\mathrm{SII}]_{671.7}$, and $[\mathrm{SII}]_{673.1}$. The spectrum of NW shows a broad H$\alpha$ line in absorption at almost 0--velocity with a weak H$\alpha$ emission line at the recession velocity of Arp~158.\label{fig:spectra}}
\end{figure*}

To gain insight into the physical nature of the selected regions and estimate their metallicity, we have also performed optical spectroscopy. Observations were carried out in service mode under good weather conditions at the William Herschel Telescope. Two long--slit spectra were obtained by our team using the ISIS (Intermediate dispersion Spectrograph and Imaging System) instrument with an exposure time of 1200~s. Each slit is 4\arcmin\ long and 2\arcsec\ wide. The first slit was aligned along the 3 bright condensations NE, NC, and NW, and the bar, with a position angle of 175.5$^\circ$. The second one is aligned along the eastern tidal arm and the FUV bright condensation capturing clumps C1, C2 and C5, with a position angle of 15.5$^\circ$. Each slit was observed using the R600R grating (dispersion of 0.50~\AA/pixel) from 612~nm to 820~nm in the red arm of ISIS and the R600B grating (dispersion of 0.44~\AA/pixel) from 360~nm to 540~nm in the blue arm of the instrument. An image of the system in the optical along with the position of the slits is presented in Fig.~\ref{fig:system-image}. Data reduction was carried out with the {\sc longslit} package of {\sc iraf} in a standard way. Flux calibration was carried out using the spectrophotometric star Feige 15 observed the same night. The emission lines were measured with {\sc splot} within the {\sc IRAF} package. We present the data in Fig.~\ref{fig:spectra}.

In Table \ref{tab:spectro} we list the fluxes of the main emission lines we detected in the blue or red spectra depending on their wavelength.

\begin{table*}[!htbp]
\caption{Spectroscopic observations. \label{tab:spectro}}
\begin{tabular}{lccccc}
\hline
Line&NE&NC&NW&C2&C5\\\hline
$[\mathrm{OII}]_{3727}$& $20.18\pm0.97$&-- &-- &-- &--\\
H$\gamma$&               $2.11\pm0.44$&-- &-- &-- &--\\
$[\mathrm{OIII}]_{4363}$&$1.11\pm0.15$&-- &-- &-- &--\\
H$\beta$&$7.83\pm0.37$                &-- &-- &-- &--\\
$[\mathrm{OIII}]_{4959}$&$2.81\pm0.39$&-- &-- &-- &--\\
$[\mathrm{OIII}]_{5007}$&$7.52\pm0.55$&-- &-- &-- &--\\
$[\mathrm{OI}]_{6300}$&$0.72\pm0.12$&$0.13\pm0.08$&-- &-- &--\\
$[\mathrm{NII}]_{6548}$&$4.79\pm0.11$&$0.57\pm0.12$&-- &$0.34\pm0.02$&$0.12\pm0.02$\\
H$\alpha$       &$49.8\pm0.15$       &$1.53\pm0.12$&$0.59\pm0.68$&$4.73\pm0.02$&$1.37\pm0.02$\\
$[\mathrm{NII}]_{6584}$&$16.7\pm0.12$&$2.03\pm0.17$&-- &$0.94\pm0.03$&$0.39\pm0.02$\\
$[\mathrm{SII}]_{6717}$&$7.42\pm0.14$&$0.84\pm0.19$&-- &$0.64\pm0.02$&--\\
$[\mathrm{SII}]_{6731}$&$5.89\pm0.17$&$0.64\pm0.16$&-- &$0.40\pm0.03$&--\\
\hline
EW(H$\alpha$)&-23&-2.3&-0.2&-174&-44\\
\hline
v&5018&4785&--&4513&4878\\
\hline
\end{tabular}
\tablefoot{Line fluxes are in $\mathrm{10^{-15}\ erg\ s^{-1}\ cm^{-2}\ \AA^{-1}}$, the equivalent widths $\mathrm{EW(H\alpha)}$ are in \AA, the velocities v are in km~s$^{-1}$.}
\end{table*}

\section{Results}
\label{sec:results}

\subsection{Optical and near--infrared morphology}

The optical image of the system is presented in the top panel in Fig.~\ref{fig:system-image}. Several features can be noted. First of all, as we mentioned earlier, there are 3 almost aligned bright concentrations, NE, NC and NW. At least 2 of them are galactic nuclei. The nature of the westernmost one, NW is uncertain. We will examine it further in Sec.~\ref{ssec:metallicity}. North of this alignment, we can see that there is a prominent dust lane. In the same alignment, we can see a tidal tail extending to the West. The smooth change of the brightness of the tail starting from the interacting galaxies suggests that it contains significant quantities of stars stripped from the parent galaxies which is confirmed by the fact that the tail is clearly seen in the near--infrared in 2MASS images. Another, more prominent, tidal tail is visible that extends from C4 to the East--South--East. This tail seems to be in the process of being detached from the parent galaxies as the drop in the brightness towards the centre of the tail suggests. In addition, it appears to be much fainter in the near--infrared compared to the northern one, probably because of the presence of fewer evolved stars in this tail. Conversely, its tip is much brighter and contains a number of blue knots suggesting on--going star formation. 

\subsection{Metallicity of the clumps}
\label{ssec:metallicity}

Spectroscopic observations of the clumps allow us to ascertain their nature and determine the oxygen abundance for regions photo--ionised by massive stars. All 3 nuclei NE, NC, and NW exhibit some H$\alpha$ emission, however the equivalent width is much larger in the case of NE. Examining the line ratios for each nucleus in the BPT diagram \citep{baldwin1981a}, it shows that NE is most likely a nuclear starburst. As for NC it suggests that it is an AGN, either an extreme LINER or a Seyfert. Finally, even though the H$\alpha$ line is detected at the redshift of the system for NW the spectrum is typical of a star at zero velocity with a strong Balmer line in absorption, confirming the suggestion of \cite{chincarini1973a} that this compact object is a star in the Milky Way rather than a galactic nucleus. Upon close examination this is consistent with the morphology in the near--infrared images, with NW that seems more compact than NE or NC.

We calculate the oxygen abundance from strong emission lines. As the emission lines were not detected in the blue part of the spectrum except in the case of the NE clump (Fig.~\ref{fig:spectra}) we rely on the NII@658.4~nm to H$\alpha$ ratio \citep{denicolo2002a} which is little affected by the extinction due to the wavelength proximity of the two lines. This estimator is of course only valid for regions photo--ionised by massive stars and cannot be used in the case of AGN, which excludes region NC.

The oxygen abundances we estimate for photo--ionised regions are close to the solar one. In terms of $\mathrm{12+\log O/H}$, the eastern nucleus NE has $8.77\pm0.07$, C2 has $8.61\pm0.09$ and C5 has $8.72\pm0.07$. The 1--$\sigma$ error bars are driven by the uncertainty on the N2 relation. Such a high metallicity for the tidal features is typical of this type of objects as gas is stripped from the radially mixed \citep{kewley2010a} metal--rich parent galaxies.

\subsection{Star formation rates and extinction\label{ssec:SFR}}

We determine the SFR both from the FUV emission and from the dust luminosity at 24~$\mu$m. The former one is sensitive to the photospheric emission of massive stars over a timescale of $100\times10^6$~years but is affected by internal extinction. The latter one is due to dust heated by massive stars but only probes star formation that is affected by extinction and can be contaminated by dust heating due to older stellar populations.

As we can see in Fig.~\ref{fig:system-image}, the morphology of the system as seen in star formation tracers is remarkably different from what can be seen in the optical bands. First of all, NE exhibits a particularly strong emission at 24~$\mu$m, which is consistent with the presence of a nuclear starburst. Emission at 24~$\mu$m can be seen throughout the rest of the interacting galaxies without any clear peak. Conversely, there is an UV peak for clump C3 with no corresponding peak at 24~$\mu$m. The tip of the westernmost tail, C2, also shows a strong peak at 24~$\mu$m with a weaker UV counterpart. In contrast C5 shows a strong UV peak with a weaker counterpart at 24~$\mu$m.

To estimate the SFR from the respective luminosities in those bands, we use: $\mathrm{SFR(FUV)=9.5\times10^{-22}L_\nu(FUV)}$, with $\mathrm{L_\nu}$ the luminosity density in $\mathrm{W~Hz^{-1}}$ \citep{kennicutt1998a}, and $\mathrm{SFR(24)=1.83\times\left[L(24)/10^{36}\right]^{0.83}}$, where $\mathrm{L(24)}$ is in W and is defined as $\mathrm{\nu L_\nu}$ at 24~$\mu$m \citep{relano2007a} which is adapted for individual star--forming regions and provides results typically within 0.2~dex of other similar estimators \citep{calzetti2010a}. These estimators have been converted to a \cite{kroupa2001a} IMF and assume a constant star formation rate over 100~Myr following \cite{calzetti2007a}. As those bands are biased respectively against and towards extinction, we also estimate the SFR from the combination of these bands using \cite{leroy2008a} relation: $\mathrm{SFR(24+FUV)=6.8\times10^{-22}L_\nu(FUV)+2.67\times10^{-23}L_\nu(24)}$, with $\mathrm{L_\nu}$ in $\mathrm{W\ Hz^{-1}}$.

Our spectroscopic observations do not allow us to correct for the extinction except in region NE. We have therefore decided to use the relation provided by \cite{buat2005a} to estimate the extinction in FUV from the ratio between the total infrared (TIR) emission and the FUV emission. Even though their relation was determined using galaxies rather than subregions in galaxies, the influence of the diffuse emission should remain minimal. Indeed the \cite{buat2005a} sample is mostly consistuted of late--type galaxies for which the diffuse emission is small \citep{sauvage1992a}. As we have only one infrared band, we use the following relation to estimate the TIR luminosity from the 24~$\mu$m luminosity provided for galaxy subregions by \cite{boquien2010a}: $\mathrm{L(TIR)=9.14\times10^4L(24)^{0.887}}$ with L(24) and L(TIR) in W. The extinction Av is determined from A$\mathrm{_{FUV}}$ assuming a \cite{calzetti2001a} attenuation curve. The SFR and estimates of the attenuation are presented in Table~\ref{tab:SFR}. These results are uncertain for the NC region due to the contamination by an active nucleus.

\begin{table*}[!htbp]
\caption{SFR and extinction.\label{tab:SFR}}
\begin{tabular}{llllll}
\hline
Region & 24~$\mu$m & FUV & FUV (corrected) & 24+FUV&Av\\
&(M$_\odot$~yr$^{-1}$)&M$_\odot$~yr$^{-1}$&M$_\odot$~yr$^{-1}$&M$_\odot$~yr$^{-1}$&mag\\
\hline
NC&$0.73\pm0.02$&$0.03\pm0.002$&$0.65\pm0.04$&$0.73\pm0.03$&1.3\\
NE&$1.52\pm0.05$&$0.04\pm0.002$&$1.32\pm0.07$&$1.75\pm0.07$&1.4\\
C2&$0.13\pm0.01$&$0.02\pm0.001$&$0.14\pm0.01$&$0.10\pm0.01$&0.8\\
C3&$0.29\pm0.01$&$0.06\pm0.003$&$0.34\pm0.02$&$0.27\pm0.01$&0.8\\
C5&$0.05\pm0.007$&$0.01\pm0.001$&$0.06\pm0.004$&$0.04\pm0.004$&0.6\\
\hline
\end{tabular}
\tablefoot{SFR in units of M$_\odot$~yr$^{-1}$ in the UV and in the IR in the 5 CO pointing beams. The extinction is in magnitude. The uncertainties only take into account the errors on the fluxes.}
\end{table*}

The uncorrected FUV yields a much lower SFR than 24~$\mu$m. This indicates that all regions suffer from strong extinction. This is particularly the case in strongly star--forming regions such as NE which has $\mathrm{A_{V}=1.4}$~mag. The Balmer decrement for this region gives an extinction of approximatively $\mathrm{A_{V}=2.1}$~mag. This is expected as the Balmer decrement measures the extinction of the gas. The relation of \cite{buat2005a} measures the extinction of the continuum, which is smaller \citep{calzetti1997a}. More quiescent regions in the system have a smaller extinction, down to $\mathrm{A_{V}=0.6}$~mag for C5. The combination of the FUV and 24~$\mu$m luminosities yields SFR that are higher than the FUV corrected ones for the 2 most extinguished regions and smaller than the FUV corrected ones for the others but no more than a factor of 1.5. IRAS (Infrared Astronomical Satellite) observations at 60~$\mu$m and 100~$\mu$m are a better estimator of the TIR as they trace the dust in thermal equilibrium with the radiation field at the peak of the emission. The total flux of the system is 0.24~Jy at 25~$\mu$m, assuming $F_\nu(24)=F_\nu(25)$, 2.03~Jy at 60~$\mu$m and 4.51~Jy at 100~$\mu$m. Estimating the TIR using equation 5 from \cite{dale2002a} we find $\mathrm{3.1\times10^{10}~L_\odot}$ which yields $\mathrm{SFR(IRAS)=3.74~M_\odot~yr^{-1}}$ using equation 4 from \cite{kennicutt1998a} converted to a \cite{kroupa2001a} IMF. The total SFR that we derived from 24+FUV is $\mathrm{2.89~M_\odot~yr^{-1}}$. This is only slightly lower than the IRAS value, most likely due to missing extended emission.

\subsection{Gas properties}

\subsubsection{CO lines}

CO(1--0) and CO(2--1) emission was detected for all pointings except for C2 for which there is a marginal detection in CO(1--0) and no detection in CO(2--1). The quality for the other positions is sufficient to determine the CO(2--1)/CO(1--0) ratios. The line ratios in NC (0.9$\pm$0.1) and C3  (0.9$\pm$0.1) are consistent with the average value found in nearby galaxies \citep[0.89,][]{braine1992a}, if the CO emission comes from a region roughly as large as the CO(1--0) beam. At these positions, optically thin emission, for which a line ratio $>1$ would be expected, can be excluded. Indeed, at the local thermodynamical equilibrium for a high temperature and high opacity we have CO(2--1)/CO(1--0)=1. In the case of optically thin emission we expect line ratios $>1$ with an asymptotic case of CO(2--1)/CO(1--0) = $\left(\nu_{21}/\nu_{10}\right)^2$ for optically thin gas in local thermodynamical equilibrium with a sufficiently high temperature \citep{wilson2009a}. Both the NE clump and C5 exhibit line ratios above 1 ($1.3\pm0.2$ for the NE clump, and $1.5\pm0.6$ for C5). This could be due to (i) optically thin emission or (ii) emission concentrated to a region smaller than the CO(1-0) beam. Since both the 24~$\mu$m and the CO emission are strong at these positions, we think that option (ii) is more likely. Note however that the beam sizes are different which makes strong conclusions difficult.

\subsubsection{Gas masses}

We calculate the molecular masses using equation 1 from \cite{braine2001a}: 

\begin{equation}
 \mathrm{M_{H_2}=I_{CO}X_{CO}D^2\Omega2m_p},
\end{equation}
where $\mathrm{I_{CO}}$ is the observed line intensity in the beam in units of K~km~s$^{-1}$, $\mathrm{X_{CO}}$ is the conversion factor to estimate the molecular mass from CO taken as $\mathrm{2\times10^{20}~cm^{-2}~(K~km~s^{-1})^{-1}}$, D the distance of the system, $\Omega$ the solid angle of the beam ($\mathrm{1.13\theta^2}$, $\theta$ being the beam FWHM), and $\mathrm{m_p}$ the mass of a proton. Similarly to \cite{bigiel2008a}, we do not take into account the helium mass to compute the molecular mass. It corresponds to $\mathrm{M_{H_2}=75I_{CO}D^2\Omega}$, with $\mathrm{M_{H_2}}$ in $\mathrm{M_\odot}$, $\mathrm{I_{CO}}$ in K~km~s$^{-1}$, D in Mpc, and $\Omega$ in arcsec$^2$.

The low resolution of HI data makes it difficult to determine the HI mass within the CO(1--0) beam, especially in the main bodies of the galaxies. As C5 is relatively well isolated from the rest of the interacting system it is easier to measure its HI mass. However, rather than examining the mass encompassed in the CO(1--0) beam, we look at the column density in the pixel corresponding to the centre of the CO(1--0) beam in order to limit the blending with the HI emission from nearby regions. The derived molecular gas mass as well as molecular and atomic gas column densities are presented in Table~\ref{tab:masses}.

\begin{table*}[!htbp]
\caption{Atomic, molecular gas mass and column density in the 5 CO pointing beams.\label{tab:masses}}
\begin{tabular}{cccccccc}
\hline
Region & M(H$_2$) & N(H$_2$) & N(HI) & 2N(H$_2$)/N(HI)&depletion timescale\\
&($10^7$~M$_\odot$)&($10^{19}$~cm$^{-2}$)&($10^{20}$~cm$^{-2}$)&&($10^8$~yr)\\
\hline
NC&$155.1\pm16.9$&$201.6\pm21.9$&$13.5\pm1.4$&$2.99\pm0.44$&21.2\\
NE&$90.3\pm10.8$&$117.4\pm14.0$&$8.6\pm0.9$&$2.73\pm0.42$&5.2\\
C2&$5.8\pm1.5$&$7.6\pm2.0$&$10.9\pm1.1$&$0.14\pm0.04$&5.8\\
C3&$55.4\pm6.4$&$72.0\pm8.2$&$14.0\pm1.4$&$1.03\pm0.16$&20.5\\
C5&$6.8\pm1.0$&$8.8\pm1.3$&$7.1\pm0.7$&$0.25\pm0.04$&17\\
\hline
\end{tabular}
\end{table*}

The system contains large amounts of molecular gas, particularly in the nuclei. However, the CO--to--H$_2$ conversion factor is uncertain in these environments and it is likely that there is more CO emission per unit H$_2$ in galactic centres. The molecular mass in the tidal features is lower, $5.8\times10^7$~M$_\odot$ in C2 and $6.8\times10^7$~M$_\odot$ in C5. Comparing to the sample of TDG candidates of \cite{braine2001a}, it is in the lower range of observed molecular masses.

Both nuclei are significantly dominated by their molecular phase, which is commonly observed in the centre of spiral galaxies \citep{bigiel2008a}. C5, on the other hand, is HI dominated and its molecular to atomic gas mass ratio is typical of what is observed in collision debris \citep{braine2001a}.

\subsection{Kinematics of the system}
\label{ssec:kinematics}

\begin{figure}[!htbp]
\center
\includegraphics[width=\columnwidth]{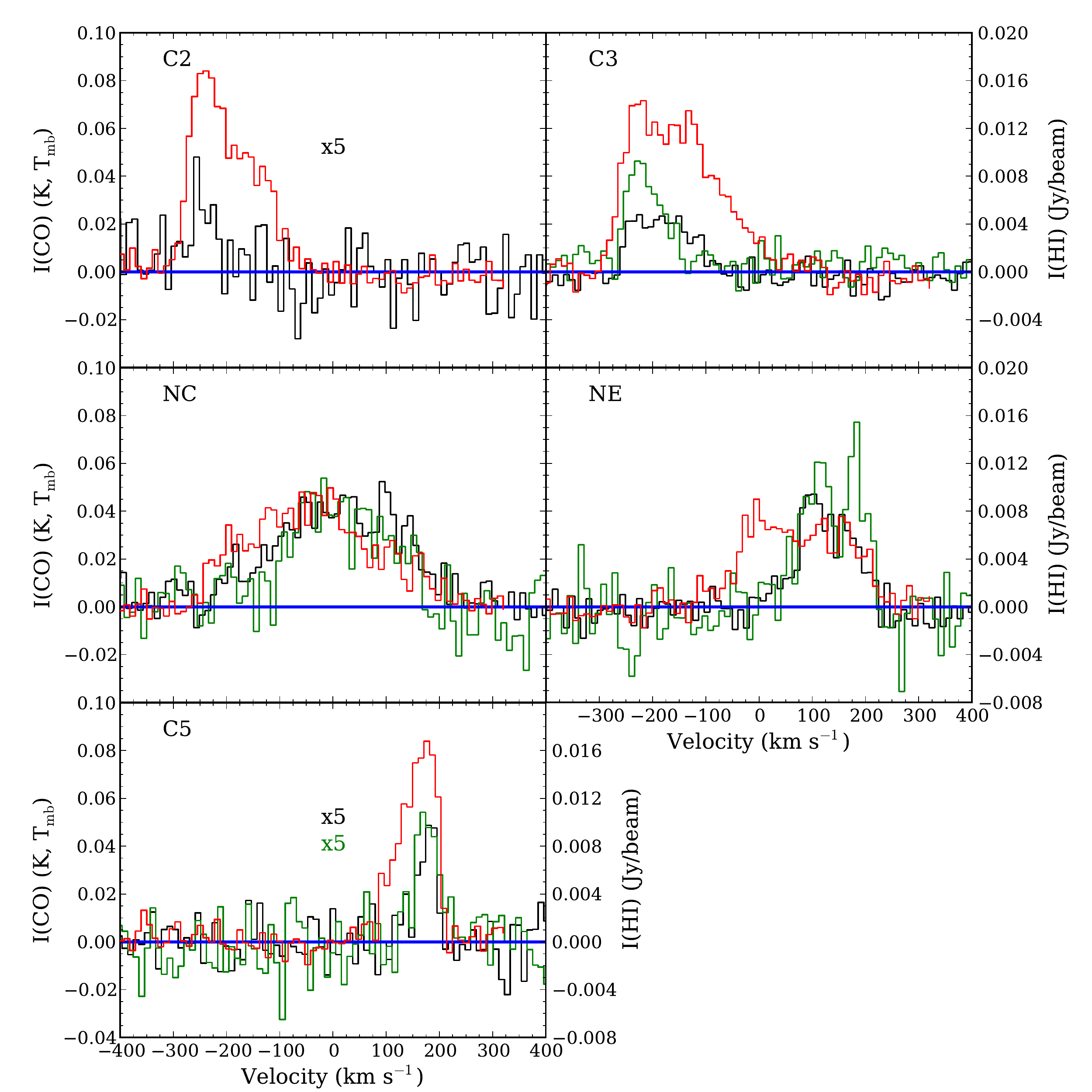}
\caption{Spectra of the five regions. CO(1--0) is shown in black, CO(2--1) in green, and HI in red. The HI spectra correspond to the central pixel in the CO beam in order to limit the blending with adjacent sources. We have multiplied the flux of the CO spectra by 5 in the case of C2 and C5 to make them more visible. All spectra have been centred to a velocity of 4800~km~s$^{-1}$.\label{fig:HI-CO-spectra}}
\end{figure}

A comparison of the HI, CO(1--0) and CO(2--1) spectra of the four CO--detected regions is presented in Fig.~\ref{fig:HI-CO-spectra}.

The recession velocity of the system that we determined through optical spectroscopy, CO and HI observations is consistent with the value of 4758~km~s$^{-1}$ mentioned in the introduction. The system rotates clockwise, the C5 and NE regions having a recession velocity comprised between 4900~km~s$^{-1}$ and 5000~km~s$^{-1}$ whereas the western C3 and C2 regions have a smaller velocity between 4500~km~s$^{-1}$ and 4700~km~s$^{-1}$.

The NE region presents a double--peaked profile centered around 4950~km~s$^{-1}$. It is most prominent in CO(2--1) in which the peaks are separated by a velocity of 70~km~s$^{-1}$. The peaks are not as strong in CO(1--0) and HI, probably due to additional contamination because of the larger beam. This may be the signature of a rotating ring of star formation in the nucleus of the galaxy with blueshifted and redshifted sides, consistent with the optical spectrum showing the presence of a nuclear starburst. Also, such a signature could be produced by a very strong molecular outflow. Finally another, although unlikely, possibility is that in the most central regions, CO is dense enough to be seen in absorption.

Regions located within the galaxy have a large CO line--width, which is likely due to the rotation of the galaxy as for NC the width of the CO(2--1) line, observed with a smaller beam, is narrower than that of the CO(1--0) line, observed with a beam twice as large. The line width reduction is minimal in the case of NE, showing the the bulk of the emission must come from the nucleus rather than from the disk of the galaxy.

HI emission displays a wider profile due to blending by nearby regions and by the fact that HI is more loosely bound compared to molecular gas. This is particularly visible in C5. The NE region shows two features. One corresponds to the double--peaked component mentioned earlier whereas the other one is at a recession velocity around 4800~km~s$^{-1}$. This probably corresponds to gas funneled to the nucleus, feeding the current starburst.

TDGs are by definition bound objects. Rotation curves have been clearly detected in a number of objects \citep{bournaud2004a,bournaud2007a}. The observations of Arp~158, unfortunately, do not have enough resolution to measure the rotation curve of C5. However, the CO line profile nevertheless provides us with an indication of the mass within the CO beam. Indeed the CO traces the densest regions which are most likely gravitationally bound, as opposed to the HI which traces lower density gas. This allows us to compute the virial mass assuming that C5 is kinematically detached from the interacting galaxies. We estimate the mass using the following relation: $\mathrm{M_{vir}=R\Delta V^2/G}$, where $\mathrm{M_{vir}}$ is the virial mass, R is half of the CO beam size, $\mathrm{\Delta V}$ is the line FWHM and G is the gravitational constant. For the C5 clump with $\mathrm{\Delta V}=40$~km~s$^{-1}$ we find $\mathrm{M_{vir}}=1.3\times10^9$~M$_\odot$. Such a virial mass is close to the molecular mass of Arp~105S for instance \citep{braine2001a}. This mass is smaller than the total HI mass in C5 according to \cite{iyer2004a}. This is not surprising as the HI is more extended. We have seen in Table \ref{tab:masses} that for C5, $\mathrm{2N(H_2)/N(HI)=0.25}$. This leads to a minimal mass $\mathrm{M(HI+H_2+He)=4.7\times10^8}$~M$_\odot$ within the CO beam, which is about 36\% of the virial mass. As the HI beam covers nearly 4 times the CO beam area, the HI column density at higher resolution could easily be several times higher or indicative of the presence of dark baryons. Indeed, the stellar mass derived from the SED (Spectral Energy Distribution) fit in Sec.~\ref{sec:discussion} is in the range of $\mathrm{0.7\times10^8~M_\odot}$ to $\mathrm{1.3\times10^8~M_\odot}$, which is too small to explain the discrepancy. New high resolution HI observations are required to answer this question.

\section{Discussion}
\label{sec:discussion}

\subsection{Nature of the system}

What is the exact nature of the clumps in the system? We have seen that we are in the presence of a 2--body merger at an intermediate stage in the \cite{toomre1977a} sequence, with the disks already merging but the 2 nuclei still clearly separated. Spectroscopic observations have shown that NE is a nuclear starburst in which the efficiency of star formation should be higher than in the rest of the system. These same spectroscopic observations have also shown that NC is an AGN, and as such this region cannot be used to discuss star formation. We examine here the SED of these clumps which provide us with information regarding their stellar content as well as their SFH (Star Formation History).

\begin{figure}[!http]
\center
\includegraphics[width=\columnwidth]{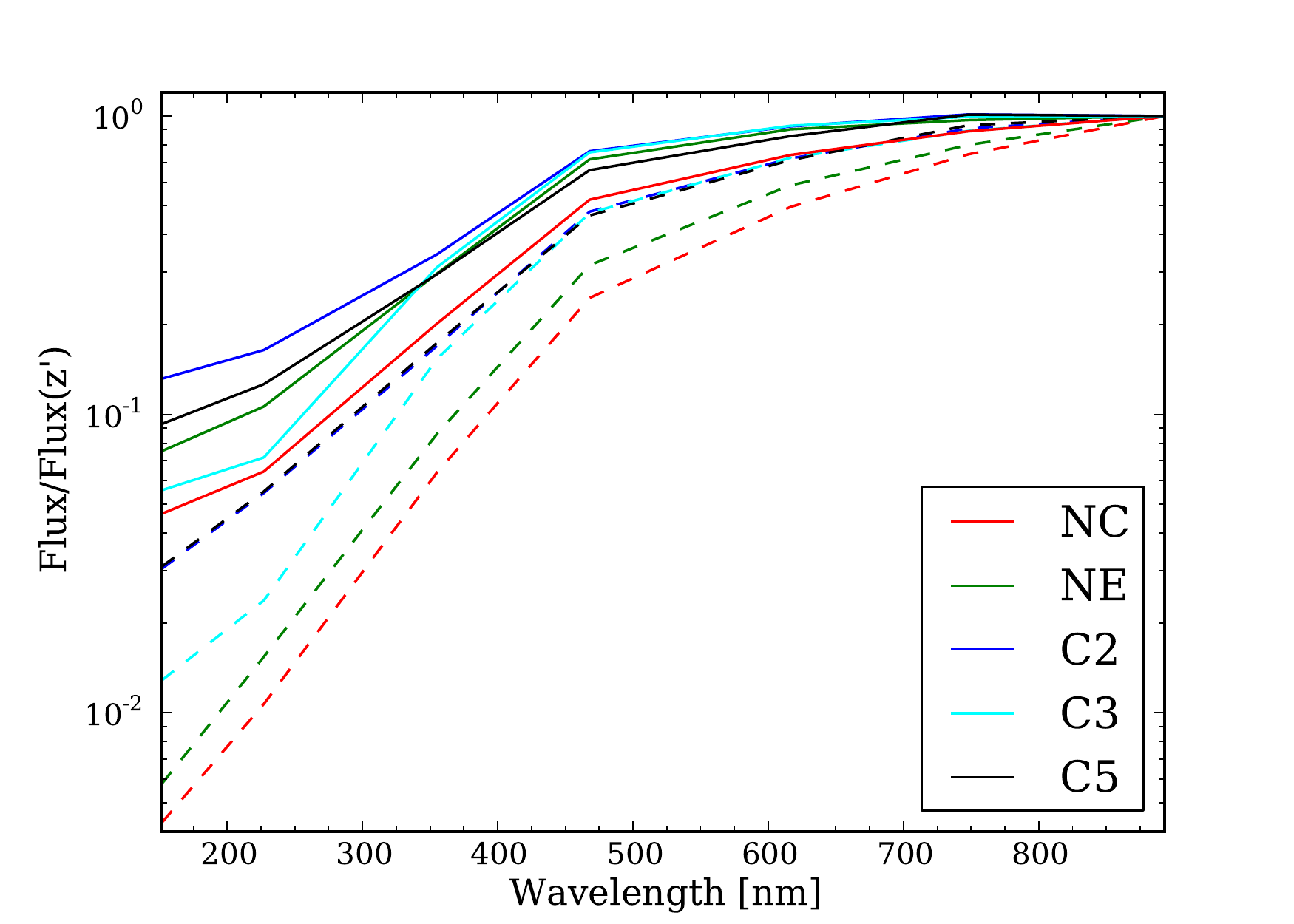}
\caption{SED of all regions normalised to the flux density in the $z'$ band. The red, green, blue, cyan and black lines represent respectively the NC, NE, C2, C3 and C5 regions. The dashed lines represent the SED not corrected for the internal extinction whereas the solid lines represent the SED corrected for the extinction assuming a starburst curve \citep{calzetti2001a} and the extinction values listed in Table \ref{tab:SFR}. Note that the extinguished SEDs of C2 and C5 are similar.\label{fig:sed}}
\end{figure}

In Fig.~\ref{fig:sed} we plot the SED of all regions normalised to the $z'$ band.
We see that the tidal features, C2 and C5, have a distinctly different SED, their UV emission being significantly higher relatively to the optical emission compared to other regions. This is most likely due to a combination of several factors. First the interaction triggered a burst of star formation that increases the UV luminosity. The second point is that those tidal features primarily come from the external regions of the galaxies which contain proportionally fewer evolved stars that emit in the optical compared to nuclear regions, this can be seen in Table \ref{tab:fluxes}. Conversely, the optical fluxes for regions in the galaxies are enhanced by the photospheric emission of the evolved stellar populations. Finally, as listed in Table~\ref{tab:SFR}, the extinction is lower in tidal features compared to more central regions. If we compare C2 and C5 to the extinguished SED, the NE region shows a relatively weak UV emission due to its high extinction enshrouding the nuclear starburst and a strong optical continuum. We see that for the extinction corrected SED, the difference is not as strong compared to the NE clump for instance, showing the importance of the extinction. Finally, NC, the likely host of an AGN, presents a weak UV luminosity with a strong optical emission from old stars.

Beyond the SED, several elements indicate that C5 is a TDG. First of all we see in Fig.~\ref{fig:system-image} that between C4 and C5 there is a faint, extended bridge connecting the two clumps showing the transfer of material from the galaxies to the tip of the tail. This bridge is most likely constituted of evolved stars stripped from the parent galaxies. Unfortunately, the resolution of the HI map does not allow us to evaluate the amount of atomic gas in the bridge. Could it be an interloping object? The optical, HI and CO spectra retrieve a similar recession velocity for C5 showing that the gas and the stars are co-incident. Moreover, the velocity is consistent with the global velocity gradient of the system. This makes it unlikely that C5 is an interloper. Quite evidently, C5 strongly resembles to dwarf galaxies with its relatively low mass, blue colour and low SFR. However, such galaxies generally have a strongly sub--solar metallicity \citep{tremonti2004a}. As we have seen in Sec.~\ref{ssec:metallicity}, its metallicity is similar to that of the C2 clump. A possibility would be that these stars actually come from a disrupted low surface brightness galaxy. The interaction could have easily disrupted the baryonic disk. However as we have shown in Sec.~\ref{ssec:kinematics}, the virial mass ($1.3\times10^9$~M$_\odot$) is about 3 times that of the visible (HI+H$_2$+He) mass. Such an interaction would not have been able to destroy the massive dark matter halo of the galaxy. This makes this scenario unlikely. While its high metallicity may be surprising as the interaction strips gas from external regions more easily, the tidal field generates an important mixing of the gas, with the torque injecting lower metallicity gas towards the nucleus, homogeneising the metallicity across the disk \citep{kewley2010a}. A higher metallicity at a given luminosity is one of the best diagnostics to separate dwarf galaxies from recycled objects \citep{duc1998a}. Using the metallicity--luminosity relation for dwarf galaxies of \cite{richer1995a}, C5 would be expected to have $12+\log O/H=8.1\pm0.1$, much lower than the observed value. Such a discrepancy is expected for collision debris. It appears reasonable that C5 is indeed a TDG rather than a pre--existing object currently undergoing an interaction with the two galaxies.

To gain additional insight, we fit the SED of C5 with an evolutionary synthesis model. It is based on the {\sc P\'egase ii} code \citep{fioc1997a}. We consider two populations of arbitrary relative mass: one representing the evolved population stripped from the parent galaxies and the other one the population formed in the TDG. Both populations have an exponentially decreasing SFH. The timescale of the old population is set so that 90\% of the original gas reservoir has been transformed into stars by the time of the onset of the formation of the younger component, that is after $12\times10^9$~years. This timescale is kept as a fixed parameter throughout the fitting procedure. The timescale of the younger population however is a free parameter. Because the dust has only been observed at 24~$\mu$m, we do not attempt to fit a dust model to these data as degeneracies would leave dust properties unconstrained. The model is described in much greater detail in \cite{boquien2010c}. Also, the lack of deep enoguh near--infrared observations does not allow for a precise discussion of the output parameters. Therefore we only use this fit to obtain a new estimate of the SFR with a SFH different than the one assumed by \cite{kennicutt1998a}. The current SFR inferred from the fit is 0.02~M$_\odot$~yr$^{-1}$, lower than the one calculated from the 24~$\mu$m and FUV luminosity, probably reflecting the difference in the assumed star formation history (constant versus exponentially decreasing).

The case of C2 is less evident due to the complexity of the interacting system. It does not seem detached yet from the parent galaxies. The metallicity is slightly lower at $8.61\pm0.09$ which is high enough to rule out a pre--existing dwarf galaxy. An additional hint is given by comparing the UV--to--optical SED of C2 and C5. As we have seen earlier in the present section, they are remarkably similar. The most likely possibility is that this region is the tip of a tidal tail. High resolution HI observations would certainly yield precious information regarding the exact nature of the C2 clump.

\subsection{Gas depletion timescale}

The gas depletion timescale, which is the time it would take to convert the molecular gas reservoir into stars at the current SFR, can vary between different objects, but it is unclear whether and how it varies within a given interacting system. \cite{braine2001a} showed that collision debris have a depletion timescale of the molecular gas similar to that of spiral galaxies. However, how this timescale changes as a function of the morphology in an interacting system is still an open question. In Table \ref{tab:SFR} we list the depletion timescale of the molecular gas in the CO(1--0) beams.

The shortest depletion timescale corresponds to the starburst in the eastern nucleus, which is also observed in other nuclear regions. This is expected as starbursts tend to have a shorter depletion timescale \citep{kennicutt1998b}. The tidal features C3 and C5 show a longer depletion timescale around $2\times10^9$~yr that is typical of what can be observed in spiral galaxies \citep{kennicutt1998b,bigiel2011a} and in collision debris in general \citep{braine2001a}. C2 however has a shorter depletion timescale. This may be due to an age effect as usual SFR estimators assume a constant SFR which causes an overestimate the current SFR if it is actually declining.

\subsection{Schmidt--Kennicutt law}

The Schmidt--Kennicutt law links the gas surface density to the SFR surface density: $\mathrm{\Sigma_{SFR}\propto\Sigma_{gas}^N}$. Whether and how this law varies is subject of an on--going debate in the literature \citep{kennicutt1998b,gao2004a,kennicutt2007a,leroy2008a,bigiel2008a,blanc2009a,rahman2011a,liu2011a}. Deviations between quiescent star-forming galaxies and interacting systems can have important implications regarding our understanding of galaxy formation and evolution. The fundamental reason is that the ISM (interstellar medium) of high redshift galaxies is turbulent \citep{forster2006a}, and that they are gas rich \citep{tacconi2010a,daddi2010a}. Nearby interacting systems, by having an enhanced turbulence, can be seen as analogues of high redshift galaxies. Recent high resolution simulations by \cite{teyssier2010a} show that interacting galaxies deviate from the standard Schmidt--Kennicutt law seen in spiral disks, which could be explained by the effect of gas turbulence and fragmentation. Whether star formation proceeds similarly to the Schmidt--Kennicutt law in local interacting galaxies is therefore important to gain insight into the mode of star formation at high redshift.

In a first step we examine how the molecular hydrogen column density and the SFR surface density in the different regions in Arp~158 compare to the relation derived by \cite{bigiel2008a} in the case of spiral galaxies. The SFR in Arp~158 corresponds to the 24+FUV measurement in Table \ref{tab:SFR} which is the same to that used by \cite{bigiel2008a}.

\begin{figure}[!htbp]
\center
\includegraphics[width=\columnwidth]{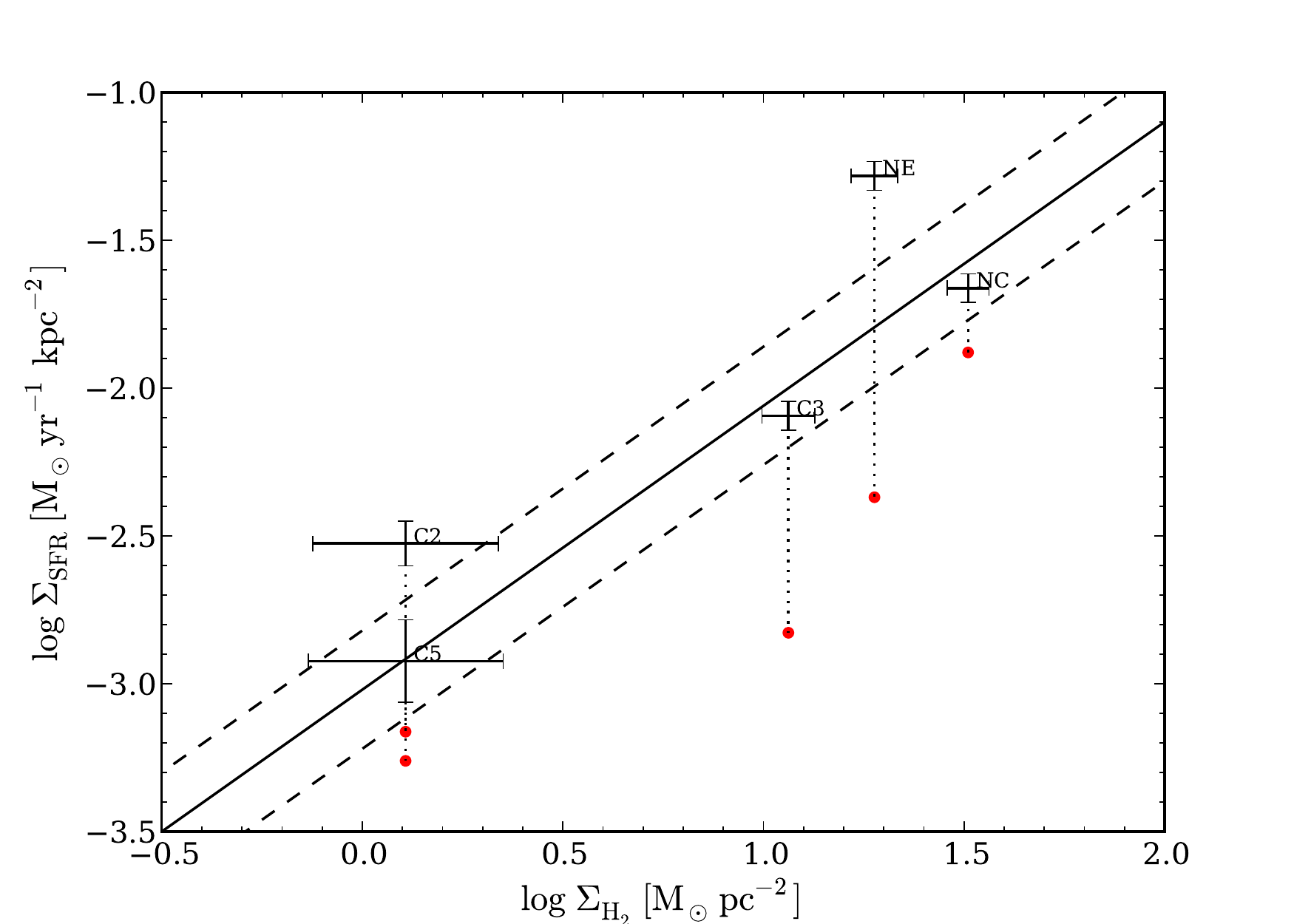}
\caption{SFR surface density versus the molecular hydrogen surface density. The points with the 3--$\sigma$ error bars represent pointings in the Arp~158 system with the SFR derived from the FUV+24~$\mu$m relation, and the red circles adopt the SFR derived from the fit of the SED assuming the burst has the extinction calculated in Table \ref{tab:SFR}. The solid black line represents the Schmidt--Kennicutt law for molecular hydrogen derived by \cite{bigiel2008a} for spiral galaxies: $\mathrm{\log\ \Sigma_{SFR}=-3.02+0.96\log\ \Sigma_{H_2}}$, with $\mathrm{\Sigma_{SFR}}$ in $\mathrm{M_\odot~yr^{-1}~kpc^{-2}}$ and $\mathrm{\Sigma_{H_2}}$ in $\mathrm{M_\odot~pc^{-2}}$. The dashed lines represent the typical 0.2~dex scatter. The molecular mass does not take into account the helium contribution in any case.\label{fig:KS-plot}}
\end{figure}

Even though there are only a few measurements, the SFR and the H$_2$ surface densities span more than one order of magnitude and are well correlated with each other in Arp~158. The NC, C3 and C5 regions follow the same relation as spiral galaxies, which is consistent with the results found by \cite{braine2001a} on TDG. The NE and C2 regions exhibit clear excesses of their SFR in comparison to their molecular gas surface density. This is easily explainable for NE as it is a nuclear starburst. This finding confirms observational and theoretical results for starburst galaxies. Another possibility for NE and C2 is that the molecular gas is more concentrated than star formation so that averaging over the beam would underestimate the molecular gas surface density. However, if the gas and star formation are equally extended it would move the points mostly parallel to the \cite{bigiel2008a} relation. Interferometric observations of the entire system would be required to answer this question. Another possibility is that a strong burst quickly depleted the molecular gas reservoir or that star formation tracers give a significantly overevaluated SFR. In all cases, if the burst SFH is decreasing, the standard SFR estimators which assume a constant SFR over 100~Myr will likely overestimate the actual SFR. As star--formation in collision debris tends to be more bursty compared to star--formation averaged over a galactic disk, this could artificially enhance the derived SFR in these regions. When using the SFR derived from the SED fitting we notice that NE and C2 are not strongly deviant anymore compared to the other regions. This shows that great caution must be employed to estimate the SFR as it can influence the results significantly, especially in the case of interacting systems in which the actual SFR can vary rapidly.

\begin{figure}[!htbp]
\center
\includegraphics[width=\columnwidth]{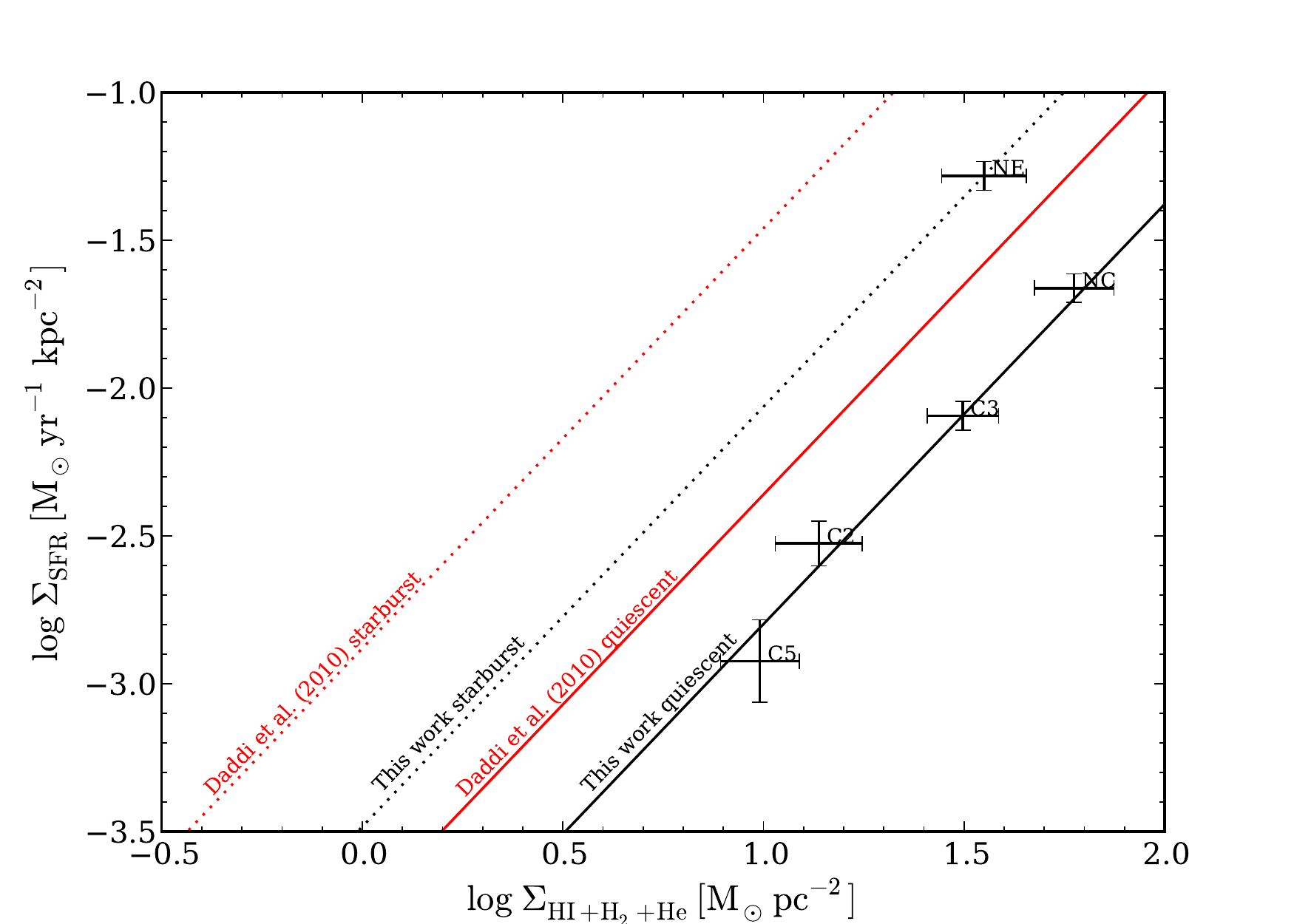}
\caption{SFR surface density versus the gas surface density, including HI, H$_2$ and He. The points with the 3--$\sigma$ error bars represent pointings in the Arp~158 system. To take He into account we have multiplied the H$_2$ column density by a factor 1.38. As the HI resolution is significantly worse than the CO beam size, we have used the column density in the central pixel as listed in Table \ref{tab:masses}. The actual column density remains uncertain and requires higher resolution HI observations. The solid red line represents \cite{daddi2010a} relation derived for BzK galaxies: $\mathrm{\log\ \Sigma_{SFR}=-3.83+1.42\log\ \Sigma_{gas}}$, with $\mathrm{\Sigma_{SFR}}$ in $\mathrm{M_\odot~yr^{-1}~kpc^{-2}}$ and $\mathrm{\Sigma_{gas}}$ in $\mathrm{M_\odot~pc^{-2}}$. The dotted red line is the same relation offset by 0.9 dex fitting ULIRGs. The \cite{daddi2010a} relation has been converted from a \cite{chabrier2003b} IMF to a \cite{kroupa2001a} one for easier comparison. The solid black line represents the best fit for the Arp~158 apertures excluding NE with the same slope as the \cite{daddi2010a} relation. Finally the dotted black line represents the same relation offset to pass through the NE region.\label{fig:KS-plot-D10}}
\end{figure}

As mentioned earlier, \cite{daddi2010a} found that starburst galaxies follow a different Schmidt--Kennicutt law than more quiescent galaxies. The interaction in Arp~158 increases the turbulence in the system. The question is whether different regions in the system also follow different relations. In Fig. \ref{fig:KS-plot-D10}, we compare the regions in Arp~158 with the relations found by \cite{daddi2010a}.

We see that similarly to what \cite{daddi2010a} found, we are seeing 2 different regimes of star formation in Arp~158, provided the SFR estimator is accurate. The first one regroups all regions, except for NE, which are well fitted by a power law with a slope of 1.42. Conversely NE presents a much higher SFR surface density for a similar surface density, with an offset which is qualitatively similar to the one found by \cite{daddi2010a} for starburst galaxies. The offset is slightly larger in the case of \cite{daddi2010a}, probably as the objects they studied are more extreme than ours. Another important point is that contrary to \cite{daddi2010a} we find this offset while keeping  the X$_\mathrm{CO}$ conversion factor constant. Using a smaller conversion factor similar to that used for LIRGs and ULIRGs would only increase the discrepancy. What really sets NE apart is not the gas surface density but the high SFR surface density. A possible explanation is that this region is not simply an inflow--driven starburst but that the increased turbulence and a fragmentation into dense clouds strongly increase the SFR surface density for the same gas surface density \citep{teyssier2010a,bournaud2010b,bournaud2011a}. An observational signature of this would be an an excess of the dense gas fraction as observed by \cite{juneau2009a}. In order to determine whether the dense gas fraction is higher, some HCN observations are required. We stress that the use of a lower X$_\mathrm{CO}$ factor, as is used for ULIRGs for instance, for the NE region would only exacerbate the discrepancy.

The presence of these 2 modes seen in a resolved way in an interacting system shows that its origin does not depend on the global mass or size of the system but that it is rather linked to the physics of the ISM at scales no larger than 1~kpc. Indeed, this scale corresponds to the largest gravitational instabilities in the ISM. The Jeans length which is of the order of 100--200~pc in nearby spirals increases up to 500--1000~pc in mergers because of higher densities and velocity dispersions. In addition this scale also corresponds to the injection scale of turbulence in the ISM \citep{elmegreen2003a,bournaud2010a}.

\section{Summary and conclusion}

In this paper we have studied how properties of star--forming regions vary across an interacting system. To do so we have combined an extensive set of archival and proprietary data tracing the molecular gas (CO), the atomic gas (HI), star formation (FUV and 24~$\mu$m), and the stellar populations. The interacting system shows a complex morphology, the disks of the 2 colliding galaxies having already interpenetrated. To ascertain the exact nature of the different regions in the interacting system we have also obtained optical spectra. In particular we have obtained a firm identification of the nuclei of the merging galaxies, which was still under debate. One, to the East, exhibits a starburst, the other one hosts an AGN. A third nucleus, to the West, turns out to be a foreground star. A brief description of the regions of interest in Arp~158 is provided hereafter.
\begin{itemize}
 \item {\it NE}: The emission of the eastern nucleus is peaked and is strong at 24~$\mu$m, suggesting an AGN or a nuclear starburst. The emission line ratios from the BPT diagram \citep{baldwin1981a} show that it is most likely the latter. The starburst is probably due to gas losing angular momentum as a result of the interaction, and inflowing towards the centre of the galaxy as a consequence.
 \item {\it NC}: Emission line ratios of the central nucleus hint at an AGN, either an extreme LINER or a Seyfert. This unfortunately compromises any accurate measure of the SFR in this regions.
 \item {\it NW}: Optical spectroscopy clearly shows that this is a foreground star with a strong H$\alpha$ Balmer line in absorption at zero velocity. A faint H$\alpha$ emission at the redshift of the system is also observed showing that the star may mask a star--forming region. A close inspection of 2MASS near--infrared images also shows that NW is qualitatively more compact than NC and NE. This confirms what \cite{chincarini1973a} found and closes the question of the nature of this object.
 \item {\it C1}: A peak is visible at 24~$\mu$m without counterpart in ultraviolet, hinting that it may be a young, dusty star--forming region that has yet to break the surrounding dust cloud. The CO emission of this clump is encompassed by the CO(1--0) beam for the NC pointing.
 \item {\it C2}: The western tip of the system is peaked at 24~$\mu$m with a faint counterpart in the ultraviolet. CO observations show that it is only marginally detected in CO(1--0) and not detected in CO(2--1), with a detection threshold at 3--$\sigma$.
 \item {\it C3}: The region is particularly peaked in FUV and NUV but does not present a similar peak at 24~$\mu$m. However CO emission is clearly detected.
 \item {\it C4}: A faint 24~$\mu$m peak can be seen without any counterpart in the ultraviolet showing the possible presence of embedded star--forming regions. The NE CO pointing overlaps partially with this region in CO(1--0).
 \item {\it C5}: The TDG candidate is clearly peaked in both the 24~$\mu$m and the ultraviolet bands. A faint, diffuse UV emission can be seen in the bridge linking it to the main bodies of the interaction. CO observations show a clear detection both in CO(1--0) and in CO(2--1).
\end{itemize}

Studying the Schmidt--Kennicutt law, most regions in the system follow closely the relation found in spiral galaxies by \cite{bigiel2008a}, with the exception of the nuclear starburst and the tip of one of the tidal tails which show a significantly larger efficiency. Comparisons with the relations derived by \cite{daddi2010a} exhibit the presence of 2 star formation modes. The detection of these 2 modes in a resolved way hints that their origin is linked to the physics of the interstellar medium at scales no larger than 1~kpc, the size of the largest gravitational instabilities and the injection scale of turbulence.
\label{sec:conclusion}

\begin{acknowledgements}
We would like to thank the anonymous referee for his/her useful comments that have helped improve this paper.

We would like to thank Caroline Simpson for kindly providing the HI data for the Arp 158 system. We thank the William Herschel Telescope staff for carrying out the spectroscopic observations in service mode. We also thank the IRAM staff for carrying out part of the CO observations in service mode.

UL acknowledges financial support from the research grants AYA2007-67625-C02-02 from the Spanish Ministerio de Ciencia y Educaci\'on and FQM-0108 from the Junta de Andaluc\'\i a (Spain). UL  warmly thanks IPAC (Caltech), where most of this work was done during a sabbatical stay, for their hospitality.

VC would like to acknowledge partial support from the EU grant FP7-REGPOT 206469.

This research has made use of the NASA/IPAC Extragalactic Database (NED) which is operated by the Jet Propulsion Laboratory, California Institute of Technology, under contract with the National Aeronautics and Space Administration.
\end{acknowledgements}

\bibliographystyle{aa}
\bibliography{article}

\end{document}